# Self-Supervised Speech Denoising Using Only Noisy Audio Signals

Jiasong Wu[a,b,c,e,*], Qingchun Li[a,b,c], Guanyu Yang[a,b,c], Lei Li[d], Lotfi Senhadji[c,e], Huazhong Shu[a,b,c]

[a]LIST, Key Laboratory of Computer Network and Information Integration, Southeast University, Ministry of Education, Nanjing 210096, China
[b]Jiangsu Provincial Joint International Research Laboratory of Medical Information Processing, Southeast University, Nanjing 210096, China
[c]Centre de Recherche en Information Biomédicale Sino-français (CRIBs), Rennes F-35042, France
[d]The 28th Research Institute of China Electronics Technology Group Corporation, Nanjing 210007, China
[e]Univ Rennes, INSERM, LTSI-UMR 1099, Rennes F-35042, France

ABSTRACT

In traditional speech denoising tasks, clean audio signals are often used as the training target, but absolutely clean signals are collected from expensive recording equipment or in studios with the strict environments. To overcome this drawback, we propose an end-to-end self-supervised speech denoising training scheme using only noisy audio signals, named Only-Noisy Training (ONT), without extra training conditions. The proposed ONT strategy constructs training pairs only from each single noisy audio, and it contains two modules: training audio pairs generated module and speech denoising module. The first module adopts a random audio sub-sampler on each noisy audio to generate training pairs. The sub-sampled pairs are then fed into a novel complex-valued speech denoising module. Experimental results show that the proposed method not only eliminates the high dependence on clean targets of traditional audio denoising tasks, but also achieves on-par or better performance than other training strategies. ***Availability***—ONT is available at https://github.com/liqingchunnnn/Only-Noisy-Training

## 1. Introduction

In our daily conversation, the intelligibility of speech communication, especially through telecommunication devices, is inevitably disturbed by various noises, and methods dedicated to speech denoising have been continuously developed. At present, research on speech denoising focuses mainly on two aspects: improving the training strategy and optimizing the denoising model.

Regarding the first aspect, most traditional speech denoising methods (Baby and Verhulst, 2019; Defossez et al., 2020; Fu et al., 2021; Pascual et al., 2017; Soni et al., 2018) are carried out on supervised training networks, which use noisy audio signals as the training input and perfectly clean audio signals as the target. We refer to this supervised strategy as **Noisy-Clean Training (NCT)**. The common problem encountered by NCT is the high cost and tedious time required to collect studio-recorded clean signals. Even if we can acquire various clean speech signals, the quantity and variety of speech conditions are often limited, and the speaker characteristics are often not universal. To overcome such limitation, some researchers have attempted to train without clean targets (Alamdari et al., 2021; Kashyap et al., 2021), such as the **Noisy-Noisy Training (NNT)** strategy. This strategy uses a mixture of clean speech and noise as the training target, and uses the same clean speech and some other noise to simulate the training input. However, we can hardly obtain multiple different noises on the same speech signal in the real world. Recently, Wisdom et al. (2020) and Fujimura et al. (2021) have proposed the **Noisier-Noisy Training (NerNT)** strategy, which recovers the original noisy speech from the mixed noisier speech. Its training target is the noisy speech, and the training input is simulated by the same noisy speech mixed with extra noise.

For the second aspect i.e., that is related to the denoising model, many researchers have devoted efforts to achieve an outstanding denoising effect. Common speech denoising networks are mostly implemented in real-valued, but there are two obvious disadvantages. The real-valued networks mainly focus on estimating the magnitude of spectrogram while reusing the phase from noisy speech for reconstruction. To solve the problem, Choi et al. (2018) present a deep complex U-Net (DCUnet). Moreover, most convolutional neural networks are designed to extract the temporal features directly. In order to alleviate the effect of the limited receptive field, Wang et al. (2021) proposed a context-aware U-Net (CAUNet), stacking a real two-stage transformer module (**rTSTM**) between the real U-Net (Ronneberger et al., 2015). The two-stage transformer module (TSTM), consisting of multiple two-stage transformer blocks (TSTBs), is used to extract local and global context information.

In summary, there still exist some problems in the two research aspects mentioned above. Firstly, these training strategies are still a long way from actual application scenarios, because we may only obtain noisy audio without any additional information in real life. Secondly, the existing deep complex denoising models focus on more precise phase estimation while ignoring the context-aware modeling of long-range speech sequences.

Two questions can then be raised:

*(1) Can we solve the speech denoising problem with neither clean speech, conditional noisy speech pairs, nor any additional noise data, but only directly from the collected noisy audio signals?*

*(2) Is the performance of the speech denoising method using only noisy audio signals better than other speech denoising methods?*

In this paper, we propose the **Only-Noisy Training (ONT)** strategy, a self-supervised speech denoising approach motivated by a similar image denoising method (Neighbor2Neighbor) (Huang et al., 2021). To the best of our knowledge, ONT is the first work that solves the speech denoising problem with only noisy audio signals in audio space. In this strategy, both of the training input and the target are generated from each of the noisy signals. By designing a specific audio sub-sampler and considering regularization loss while training, ONT can achieve denoising results similar to the Noisy-Clean Training.

Compared to some other self-supervised methods (Maciejewski et al., 2021; Saito et al., 2021; Sivaraman et al., 2021, 2022; Wang et al., 2020), the unique advantage of our method is that the ONT strategy can be applied to any supervised speech denoising tasks, and this strategy does not depend on any training model. It means that any effective speech denoising model can use our ONT strategy and be trained without any clean targets or model modifications. The proposed self-supervised framework aims at training denoising networks with only single audio samples available, without any modifications to the network structure.

---

* Corresponding author. Tel: 86-25-83794249; Fax: 86-25-83792698.
  *E-mail address:* jswu@seu.edu.cn

To achieve superior training performance while using the ONT strategy, we design an effective complex-valued speech denoising network, which inserts the proposed **complex TSTM (cTSTM)** into DCUnet by modeling the correlation between the magnitude and phase information.

Our contributions can be summarized as follows: (1) We propose a novel training strategy (i.e., Only-Noisy Training), putting an end to the training need for clean audio signals and any additional dataset limitation; (2) We train a novel complex-valued speech denoising network with the proposed strategy; (3) The experimental results demonstrate that our ONT strategy performs very favorably against other strategies on the UrbanSound8K dataset and achieves on-par or better performance on the Voice Bank + Demand benchmark dataset.

## 2. Related work

### 2.1. Training Strategy

Most speech denoising tasks are based on the Noisy-Clean Training (NCT) strategy, which use clean speech signals as the training target. For example, Pascual et al. (2017) proposed a time-domain U-Net model optimized with generative adversarial networks, Soni et al. (2018) investigated a time-frequency masking-based method with a modified adversarial training method, Defossez et al. (2020) made efforts to propose a denoising model that directly operates on the raw input waveform and generates a waveform for each source and Fu et al. (2021) implemented the idea to mimic the behavior of a target evaluation function with a neural network.

As speech signals are easily contaminated by the surrounding noise, absolutely clean signals can only be obtained in a well-controlled recording environment. Therefore, collecting such signals is costly and time consuming.

To mitigate this problem, a series of speech denoising methods that do not require clean speech signals are proposed. The most typical is the Noisy-Noisy Training (NNT) strategy, which trains the network with multiple independent noisy speech signals per scene (Alamdari et al., 2021; Kashyap et al., 2021) by applying the Noise2Noise (Lehtinen et al,. 2018) technique in the audio space. Two key conditions must be met in this strategy: the median or mean of the target distribution remains the same in the presence of noise, and the trained model is used to map the noise in the input signal to another noise in the target signal. Assuming that the noise distribution is zero-mean, the network is trained to become a noise suppressor. Moreover, the Noisier-Noisy Training (NerNT) (Fujimura et al., 2021; Wisdom et al., 2020) strategy focuses on training a denoising model to predict a noisy speech signal from a noisier signal.

Additionally, some self-supervised methods also aim at achieving outstanding denoising performance, Wang et al. (2020) investigated a self-supervised learning approach for speech denoising. Its first step is to use a training set of clean speech examples to learn an appropriate representation of the clean signals in an unsupervised way. The second step is to learn a mapping from noisy speech to clean sounds with the learned representation in the first step. Moreover, Sivaraman et al. (2021) implemented a training process using noisy data of the intended test-time speaker rather than the clean voice and personalized speech denoising models for twenty different speakers. Maciejewski et al. (2021) proposed a novel loss function which implicitly identifies noise by exploiting the inseparability of the noise and minimizes the effect of separation errors. However, none of these models have strong generality. As we know, most researchers have created valuable models in supervised tasks, if there is a common strategy that can be migrated to any fully supervised approach, the contribution to the speech denoising field will obviously be significant.

### 2.2. Denoising Model

From the perspective of the denoising methods, noisy speech can be enhanced either in the time-frequency (TF) domain or only in the time-domain. The time-domain methods can be divided into two categories. The first gathers the direct regression methods (Fu et al., 2018; Stoller et al., 2018) which learn a regression function from the waveform of a noisy mixture speech to the target speech. The second is the set of methods usually designed to extract the temporal features by the convolution neural networks or recurrent neural networks, where the speech denoising task can be treated as a sequence-to-sequence mapping problem.

As another main-stream, the TF-domain methods (Narayanan and Wang, 2013; Srinivasan at al., 2006; Xu et al., 2013; Yin et al., 2020; Zhao et al., 2016) are based on the fact that detailed structures of speech can be more separable after the short-time Fourier transform (STFT) operation. Traditional methods only use the magnitude between clean speech and mixture speech, ignoring the phase information. Therefore, some researchers have focused on the outstanding denoising effect from phase information. The complex ratio mask (CRM) (Williamson et al., 2015) was proposed to directly optimize on complex values and the complex spectral mapping (CSM) was proposed to estimate the real and imaginary spectrogram of mixture speech. Both of them have full information about the speech signal and theoretically the best speech enhancement performance can be achieved. Most of the above methods are implemented in real-valued. Creatively, DCUnet (Choi et al., 2018) was proposed to deal with the complex-valued spectrogram, and trained to estimate CRM and optimize the scale-invariant source-to-noise ratio (SI-SNR) loss (Williamson et al., 2015).

In addition to focusing on optimizing the extraction of time features for speech denoising tasks, researchers also pay attention to modeling long-range speech sequences by recurrent neural networks (RNNs), such as the long short-term memory (LSTM) and gated recurrent units (GRU). Therefore, some efforts have been made by incorporating LSTM layers between the encoder and decoder for learning long-range dependencies, while ignoring the contextual information of the speech.

Thus, Luo et al. (2020) extracted contextual information from the speech by a novel dual-path network, Vaswani et al. (2017) resolved the long-dependency problem with an effective transformer neural network, Wang et al. (2021) proposed a context-aware U-Net (CAUNet) by incorporating a TSTM into the U-Net architecture.

Therefore, it is essential to propose an effective denoising method that not only considers the extraction of time features, but also focuses on the contextual information.

## 3. Motivation and Theory

### 3.1. Motivation

The motivation for this work stems from the problems within existing training strategies. Considering the effectiveness of the Neighbor2Neighbor method, where authors show that image denoising with only a single noisy image is possible, we take it into consideration in the audio case. Through the Neighbor2Neighbor method, the single image denoising problem is successfully solved by generating sub-sampled paired images with random neighbor sub-samplers as training image pairs and using the self-supervised training scheme with a regularization term.

However, when we consider the idea in the speech denoising space, there exist some additional challenges to solve:

(1) In the training image pairs generated stage, the sub-sampled images can be directly obtained from original images. However, in audio case, speech signals can be represented in many different ways. We should consider whether to sub-sample directly from the raw audio signals or spectrograms.

(2) In the image denoising stage, images are directly input into real deep networks since they are real-valued and static. However, in audio case, it is very necessary to design a complex-valued network that can extract contextual information since the raw audio signals are time series and the corresponding spectrograms are complex-valued.

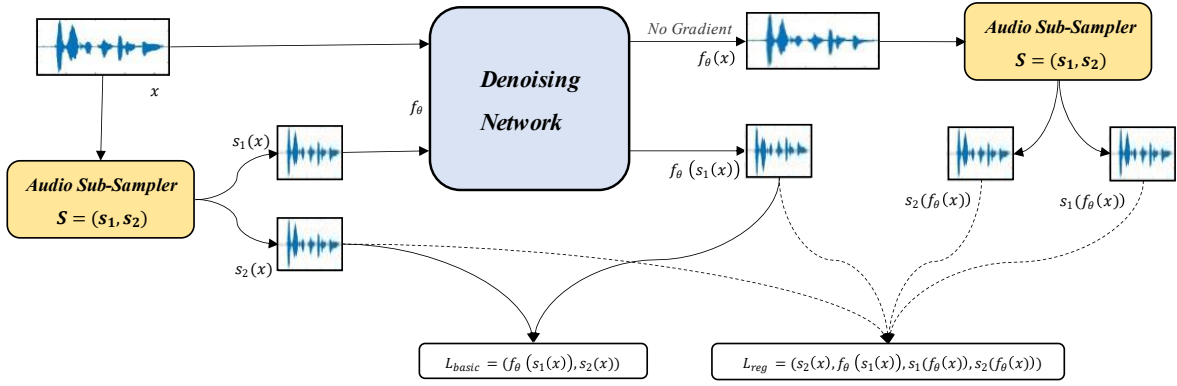

Fig. 1. Overview of our proposed ONT strategy. For a single noisy audio sample $x$, an audio sub-sampler is used to generate a pair of training samples $s_1(x)$ and $s_2(x)$, which are used as the training input and target in the denoising network. The total training loss consists of basic loss $\mathcal{L}_{\text{basic}}$ and regularization loss $\mathcal{L}_{\text{reg}}$. The basic loss is calculated from the output $f_\theta(s_1(x))$ of the denoising network and the training target $s_2(x)$, and the regularization loss requires an extra calculation of the essential difference of the ground-truth time domain values between the sub-sampled noisy audio pair.

In section 4, we will solve the above problems respectively.

*3.2. Theoretical Support*

Inspired by Neighbor2Neighbor method for the noisy image denoising work, we provide the following two key conditions for our self-supervised speech denoising strategy:

*3.2.1. Training condition*

Firstly, we consider the case of two independent noisy audio signals of similar scenes.

Suppose there is a clean audio signal $y$, and the corresponding noisy speech is $x$, such that $\mathbb{E}_{x|y}(x) = y$. When a very small signal gap $\varepsilon \neq 0$ is introduced, $y + \varepsilon$ is the clean signal corresponding to another noisy speech $z$, such that $\mathbb{E}_{z|y}(z) = y + \varepsilon$. Let the variance of $z$ be $\sigma_z^2$, then it holds as the following equation:

$$\mathbb{E}_{y,x}\|f_\theta(x) - y\|_2^2 = \mathbb{E}_{y,x,z}\|f_\theta(x) - z\|_2^2 - \sigma_z^2 \\ + 2\varepsilon \mathbb{E}_{y,x}(f_\theta(x) - y), \quad (1)$$

where $f_\theta$ is the denoising network parameterized by $\theta$, and the proof is similar to the image denoising network in Neighbor2Neighbor.

If $\varepsilon \to 0$ in (1), the noisy audio pair $(x, z)$ works as an approximation to $(x, y)$ when training the denoising network. With the small signal gap $\varepsilon \neq 0$ defined, $y$ and $z$ satisfy the condition called "similar but different". Therefore, once a suitable $y$ and $z$ satisfying this condition are found, a speech denoising network can be trained.

Next, we consider the scene of a single noisy audio signal. One possible way to construct a "similar but different" pair is to sample from adjacent but different time domain locations from the original noisy audio signal (i.e., $\varepsilon \to 0$).

Therefore, we use an audio sub-sampler $S = (s_1, s_2)$ to generate the training audio pair $(s_1, s_2)$ from the single noisy audio signal $x$. Then, the sampled audio pair can be directly used to train the denoising network in a way similar to (1), which becomes:

$$\arg\min_\theta \mathbb{E}_{x,y}\|f_\theta(s_1(x)) - s_2(x)\|^2. \quad (2)$$

Thus, the following condition can be obtained:

**Condition 2.1** *Once two paired audio signals are sub-sampled from adjacent but different time domain locations from a noisy audio signal, this pair which has similar but different characteristics can be used to train the denoising network.*

*3.2.2. Regularization Constraint*

An additional requirement needed to be considered is the different ground-truths of the sub-sampled audio pair $(s_1, s_2)$, i.e., $\mathbb{E}_{x|y}(s_2(x)) - \mathbb{E}_{x|y}(s_1(x)) \neq 0$. Therefore, the direct use of (2) is inappropriate, which will lead to over-smoothing while training. Inspired by Neighbor2Neighbor, we add a regularization term to solve the problem caused by non-zero $\varepsilon$.

Considering an ideal speech denoising network $f_\theta^*$ trained with clean audio signals, given a single noisy audio signal $x$ and a clean audio signal $y$, they satisfy:

$$\begin{aligned} f_\theta^*(x) &= y, \\ f_\theta^*(s_\ell(x)) &= s_\ell(y), \text{ for } \ell \in \{1, 2\}. \end{aligned} \quad (3)$$

Therefore, the following holds with the ideal network $f_\theta^*$:

$$\begin{aligned} &\mathbb{E}_{x|y}\{f_\theta^*(s_1(x)) - s_2(x) - (s_1(f_\theta^*(x)) - s_2(f_\theta^*(x)))\} \\ &= s_1(y) - \mathbb{E}_{x|y}\{s_2(x)\} - (s_1(y) - s_2(y)) \\ &= s_2(y) - \mathbb{E}_{x|y}\{s_2(x)\} = 0. \end{aligned} \quad (4)$$

Thus, instead of optimizing (2) directly, the following optimization problem with a constraint is considered:

$$\begin{aligned} &\min_\theta \mathbb{E}_{x|y}\|f_\theta(s_1(x)) - s_2(x)\|_2^2, \text{ s.t.} \\ &\mathbb{E}_{x|y}\{f_\theta(s_1(x)) - s_2(x) - s_1(f_\theta(x)) + s_2(f_\theta(x))\} = 0. \end{aligned} \quad (5)$$

Due to $\mathbb{E}_{y,x} = \mathbb{E}_y \mathbb{E}_{x|y}$, (5) can be rewritten into one with a regularization constraint:

$$\min_\theta \mathbb{E}_{y,x}\|f_\theta(s_1(x)) - s_2(x)\|_2^2 \\ + \gamma \mathbb{E}_{y,x}\|f_\theta(s_1(x)) - s_2(x) - s_1(f_\theta(x)) + s_2(f_\theta(x))\|_2^2, \quad (6)$$

where $x$ and $y$ denote noisy audio signals and clean audio signals respectively, $f_\theta$ represents the audio denoising network, $s_1$ and $s_2$ represent the audio sub-samplers, $\gamma$ is a tunable parameter.

Thus, the following condition can be obtained:

**Condition 2.2** *To avoid over-smoothing while training, it is necessary to consider a regularization loss by addressing the essential difference of the ground-truth time domain values between the sub-sampled pairs on the original audio.*

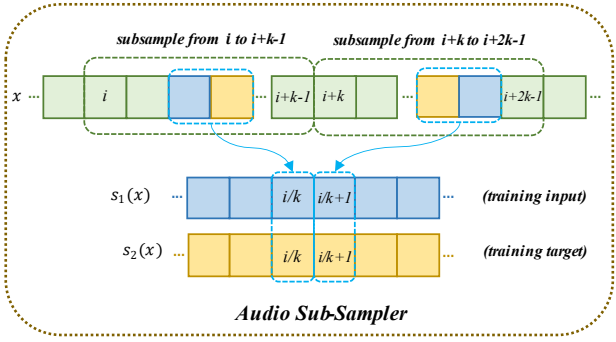

Fig. 2. The illustration of the training audio pairs generated module. The green squares represent the original noisy time domain values in turn, where $k$ is a self-set hyperparameter to control the sampling interval. From the $i$-th to $(i+k-1)$-th square, two adjacent squares are randomly selected and filled with blue and yellow respectively. The square filled with blue is considered the corresponding square of the subsampled audio $s_1(x)$, and the other square filled with yellow is considered the one of $s_2(x)$.

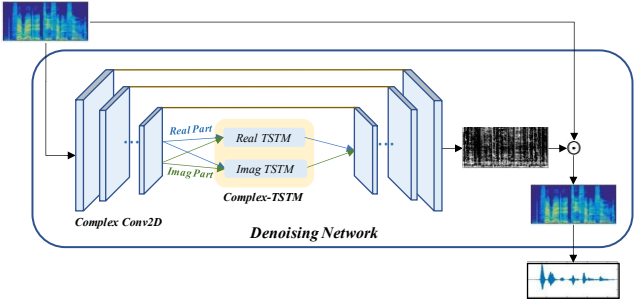

Fig. 3. The speech denoising module.

## 4. Proposed Method

In this section, we propose a novel speech denoising method trained with only noisy audio signals. The overview of our proposed ONT strategy is shown in Fig. 1. Firstly, we will introduce the detailed training conditions of the ONT strategy and the generation process of training audio pairs. Then, we will describe the details of our speech denoising network and the training loss in our strategy.

### 4.1. Only-Noisy Training (ONT) Strategy

#### 4.1.1. Application of Training Strategy

Based on the assumption in Condition 2.1, if we want to train a speech denoising network from only noisy signals, the sub-sampled audio pairs should satisfy the following requirements:

(1) Given the ground-truth $y$ of a noisy audio signal $x$, the sub-sampled audio pair $(s_1(x), s_2(x))$ has conditional independence;
(2) The underlying ground-truth gap between $s_1(x)$ and $s_2(x)$ is small.

#### 4.1.2. Generation of Training Audio Pairs

The generation process from a noisy audio signal to a training pair is shown in Fig. 2, which is mainly implemented using an audio sub-sampler. Denote a single noisy audio sample as $x$ with hyper-parameter $k$ to control the sampling interval, where $k \geq 2$. The size of the k represents the similarity between the sub-sampled pair. From the $i$-th to the $(i+k-1)$-th time-domain audio value in $x$, two adjacent random values are selected, which are used as the $i/k$-th value of the paired audio $s_1(x)$ and $s_2(x)$. Then, we shift the sampling area to the right by $k$, which then spread from the $(i+k)$-th to the $(i+2k-1)$-th value, and so on. Since $s_1(x)$ and $s_2(x)$ are selected from adjacent squares in $x$, the ground-truths of audio pairs are similar, satisfying the requirements stated in Condition 2.1 and can be used to train a speech denoising network.

Note that, we sub-sample from the raw audio but not from its spectrograms. The reason is that it is unreasonable to implement sub-sampling from spectrograms where local speech information in each time-window is extracted by STFT since different window sizes lead to different extracted speech features. In addition, sub-sampling from spectrograms will reduce the conditional independency between the two sub-sampled signals. Thus, it's inappropriate to apply sub-sampling on spectrograms during the generation process. Furthermore, a great benefit of sub-sampling from the raw audio is that any network that performs well in supervised audio denoising tasks can apply our sub-sampling method.

### 4.2. Denoising Network

The proposed speech denoising network is shown in Fig. 3, and it takes the spectrograms as input, which is derived from the sampled audio signals.

The complex encoder and decoder modules used in the denoising network are designed according to the complex module in DCUnet-10 (Choi et al., 2018). The complex 2D convolution operation (Trabelsi et al., 2018) that controls complex information flow in encoder layers contains four traditional 2D convolution operations. The complex filter $W$ is defined as $W = W_r + jW_i$, where the real matrices $W_r$ and $W_i$ represent the real and imaginary components of a complex convolution kernel. Meanwhile, the complex input matrix is defined as $X = X_r + jX_i$ so that we can get a complex output of the complex convolution operation as follows:

$$X \circledast W : F_{\text{conv}} = (X_r * W_r - X_i * W_i) + j(X_r * W_i + X_i * W_r), \quad (7)$$

where $F_{\text{conv}}$ represents the output characteristic of a complex layer.

Similar to complex convolution, we extend the real TSTM (rTSTM) to complex TSTM (cTSTM) to better model the correlation between magnitude and phase. The output $\mathbf{F}_{\text{cTSTM}}$ is defined as:

$$\mathbf{F}_{rr} = \mathbf{TSTM}_r(\mathbf{X}_r); \quad (8)$$
$$\mathbf{F}_{ir} = \mathbf{TSTM}_r(\mathbf{X}_i); \quad (9)$$
$$\mathbf{F}_{ri} = \mathbf{TSTM}_i(\mathbf{X}_r); \quad (10)$$
$$\mathbf{F}_{ii} = \mathbf{TSTM}_i(\mathbf{X}_i); \quad (11)$$
$$\mathbf{F}_{\text{cTSTM}} = (\mathbf{F}_{rr} - \mathbf{F}_{ii}) + j(\mathbf{F}_{ri} + \mathbf{F}_{ir}), \quad (12)$$

where $\mathbf{X}_r$ and $\mathbf{X}_i$ represent real and imaginary components of the complex input; $\mathbf{TSTM}_r$ and $\mathbf{TSTM}_i$ represent the real part and imaginary part of traditional TSTM; $F_{rr}$ is calculated by input $\mathbf{X}_r$ with $\mathbf{TSTM}_r$; $\mathbf{F}_{\text{cTSTM}}$ represents the complex TSTM output result.

Our final complex-valued speech denoising network is constructed by inserting a cTSTM between the encoder and decoder layers of DCUnet, ensuring that the amplitude and phase information from spectrograms can be processed and reconstructed more accurately, and the contextual information of speech will not be ignored at the same time.

### 4.3. Training Loss

In addition to satisfying the constraints of the sub-sampled audio pairs, we also need to consider the regularization loss discussed in Condition 2.2, thus the optimization problem in (6) could be considered to solve the audio denoising task successfully.

Therefore, the total loss $\mathcal{L}$ in the denoising network consists of the basic loss and the regularization loss:

$$\mathcal{L} = \mathcal{L}_{basic} + \gamma \cdot \mathcal{L}_{reg}, \quad (13)$$

where $\mathcal{L}_{basic}$ is the basic loss determined by the characteristics of denoising network, $\mathcal{L}_{reg}$ is the regularization loss used to prevent over-smoothing while training with the sub-sampled audio, and $\gamma$ is the tunable parameter introduced in (6).

*4.3.1. Basic Loss*

We combine time domain loss $\mathcal{L}_F$, frequency domain loss $\mathcal{L}_T$, and weighted SDR loss function ($\mathcal{L}_{wSDR}$) (Macartney and Weyde, 2018) to construct the basic loss. The $\mathcal{L}_{basic}$ is defined as:

$$\mathcal{L}_{basic} = (\alpha * \mathcal{L}_F + (1-\alpha) * \mathcal{L}_T) * \beta + \mathcal{L}_{wSDR}, \quad (14)$$

where $\alpha$ and $\beta$ are hyper-parameters controlling the strength of the $\mathcal{L}_F$ term and $\mathcal{L}_T$ term.

The time domain loss is computed based on the mean square error (MSE) between the enhanced waveform and the clean waveform, defined as:

$$\mathcal{L}_T = \frac{1}{N}\sum_{i=0}^{N-1}(s_i - \hat{s}_i)^2, \quad (15)$$

where $s_i$ and $\hat{s}_i$ respectively represent the $i$-th sample of clean speech and the denoised speech, and $N$ is the total number of audio samples.

The frequency domain loss can monitor the model to learn more information, resulting in higher speech intelligibility and perceived quality (Pandey and Wang, 2020). The $\mathcal{L}_F$ is defined as:

$$\mathcal{L}_F = \frac{1}{TF}\sum_{t=0}^{T-1}\sum_{f=0}^{F-1}[(|S_r(t,f)| + |S_i(t,f)|) - (|\hat{S}_r(t,f)| + |\hat{S}_i(t,f)|)], \quad (16)$$

where $S$ and $\hat{S}$ represent the clean spectrogram and the enhanced spectrogram, $r$ and $i$ are the real and imaginary parts of the complex variable, $T$ and $F$ respectively denote the number of frames and frequency bins.

The $\mathcal{L}_{wSDR}$ introduced by Lehtinen et al. (2018) is used as another term to directly optimize the well-known evaluation measures defined in the time-domain:

$$\alpha = \frac{\|y\|^2}{\|y\|^2 + \|x-y\|^2}$$
$$\mathcal{L}_{wSDR}(x,y,\hat{y}) = -\alpha \frac{\langle y,\hat{y}\rangle}{\|y\|\|\hat{y}\|} - (1-\alpha)\frac{\langle x-y, x-\hat{y}\rangle}{\|x-y\|\|x-\hat{y}\|}, \quad (17)$$

where $x$ denotes the noisy sample, $y$ denotes the target sample and $\hat{y}$ denotes the estimated output, $\alpha$ represents the energy ratio between the target speech and noise.

*4.3.2. Regularization Loss*

Given an audio pair $s_1(x)$ and $s_2(x)$ sampled from the noisy speech $x$, we use the regularization loss $\mathcal{L}_{reg}$ described in Section 2.2.2 as an additional constraint on the loss function, which is defined as:

$$\mathcal{L}_{reg} = \left\|f_\theta(s_1(x)) - s_2(x) - \left(s_1(f_\theta(x)) - s_2(f_\theta(x))\right)\right\|_2^2, \quad (18)$$

where $f_\theta$ denotes the denoising network. In order to stabilize learning, the gradient update of $s_1(f_\theta(x))$ and $s_2(f_\theta(x))$ is stopped during the training process, and the hyperparameter $\gamma$ in (13) is gradually increased to reach the best training effect.

## 5. Experiments and Results

In this section, we will provide details on our experimental verification. Starting with the experimental setup details, we will firstly verify the feasibility of the ONT strategy, then conduct comparative experiments with other training strategies and state-of-the-art methods. Additionally, we also conduct sensitivity analysis through an exhaustive series of experiments.

*5.1. Experimental Setup*

*5.1.1. Datasets*

**Evaluation of the training strategy.** Two different categories of noisy signals are used to verify the robustness of our strategy from other training strategies. While the first overlaps white gaussian noise over clean speech to generate a synthetic noisy dataset, the second overlaps different kinds of UrbanSound8K (US8K) (Salamon et al., 2014) noises over clean signals to generate real world noisy dataset. The clean speech is all from Voice Bank dataset (Valentini-Botinhao et al., 2017), including 28 speakers for the training set and 2 speakers for the testing set. In the generation process, we employ PyDub (Robert et al., 2018) to overlap the above noise over a clean audio sample, and a complete noisy speech sample is generated by truncating or repeating the noise so that it covers the whole speech segment.

**Evaluation with state-of-the-art methods.** A benchmark dataset is also exploited to compare with state-of-the-art methods. Noise and clean speech recordings are provided by the Diverse Environments Multichannel Acoustic Noise Database (DEMAND) (Thiemann et al., 2013) and the Voice Bank corpus (Veaux et al., 2013). The training set includes 11572 utterances of 28 speakers, where 8 types of noises come from DEMAND dataset (Taal et al., 2011) and 2 noises are artificially generated. By using the exact same training and test dataset split, it's easy to establish a fair comparison with other methods.

*5.1.2. Baseline Approaches*

**Evaluation of the training strategy.** Three different training strategies are evaluated as follows: **NCT** (Kashyap et al., 2021): a DCUnet model which is trained on the noisy input using the original clean speech sample as a target; **NNT** (Kashyap et al., 2021): a DCUnet model which is trained on the same inputs using the noisy targets; **NerNT** (Kashyap et al., 2021): a DCUnet model whose training target is the noisy speech, and the training input is simulated by the same noisy speech mixed with extra noise.

**Evaluation with state-of-the-art methods.** Fourteen different training strategies are evaluated as follows: **Noisy**: the original input noisy speech signal; **SEGAN** (Pascual et al., 2017): a denoising model directly operating on the raw waveform and trained to minimize the combination of adversarial and $\mathcal{L}_1$ losses; **MMSE-GAN** (Soni et al., 2018): a time-frequency masking-based method that uses a generative adversarial network (GAN) objective along with $\mathcal{L}_2$ loss; **SERGAN** (Baby and Verhulst, 2019): a denoising model training with a relativistic least-square loss for GAN training; **BLSTM (MSE)** (Weninger et al., 2015): a framework based on Long Short-Term Memory (LSTM) RNNs which is discriminatively trained according to an optimal speech reconstruction objective; **MetricGAN** (Fu et al., 2019): the first work that employs GAN to directly train the generator with respect to multiple evaluation metrics; **HiFi-GAN** (Su et al., 2020): an end-to-end feed-forward WaveNet architecture trained with multi-scale adversarial discriminators; **PHASEN** (Yin et al., 2020): a two-stream network, where amplitude stream and phase stream are dedicated to amplitude and phase prediction; **CAUNet** (Wang et al., 2021): a context-aware U-Net for speech enhancement in the time domain; **T-GSA** (Kim et al., 2020): a system for providing Gaussian weighted self-attention for speech enhancement; **DEMUCS** (Defossez et al., 2020): a novel waveform-to-waveform model, with a U-Net structure and bidirectional LSTM; **MetricGAN+** (Fu et al., 2021): an updated version of the MetricGAN framework; **NeTT** (Fujimura et al., 2021): a DNN model whose training target is the mixture of speech and noise, and training target is simulated by using the same clean speech and some other noise; **NyTT(L)** (Fujimura et al., 2021): a DNN trained to predict a noisy speech signal from a noisier signal.

*5.1.3. Evaluation metrics*

To evaluate the quality of the denoising effect, the following metrics are used: signal-to-noise ratio (SNR), segmental signal-to-noise ratio (SSNR), wide-band perceptual evaluation of speech quality (PESQ-WB) (Rix et al., 2001), narrow-band perceptual evaluation of speech quality (PESQ-NB) (Rix et al., 2001), short term objective intelligibility (STOI) (Hu and Loizou, 2007). We also adopt subjective mean opinion scores (MOSs) (Hu and Loizou, 2008) with the following three metrics: CSIG for signal distortion (from 1 to 5), COVL for overall quality evaluation (from 1 to 5) and CBAK for noise distortion evaluation (from 1 to 5).

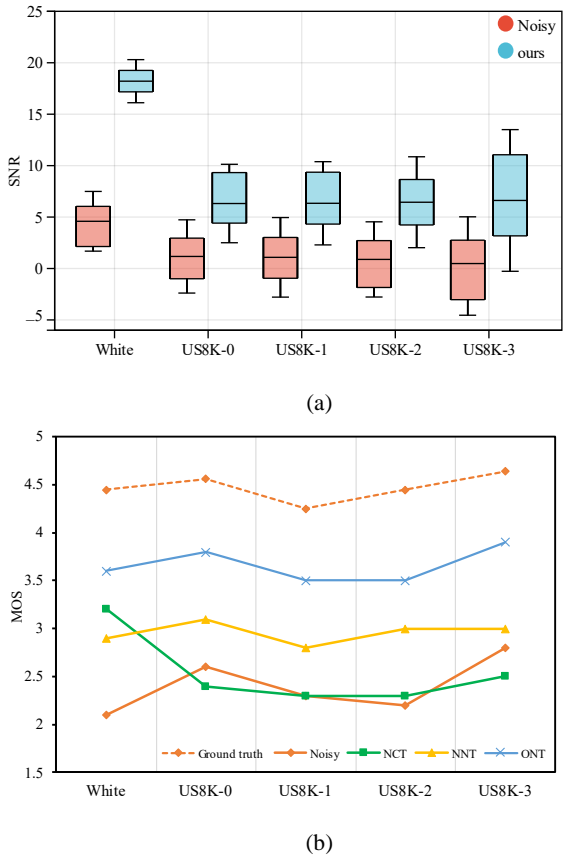

Fig. 4. Experimental results for the feasibility evaluation of our training strategy. (a) Comparison between input signals (Noisy) and ONT results. (b) The subjective evaluation results of MOS.

*5.1.4. Training details*

The original raw audio waveforms are first sampled at 48kHz. For the actual model input, complex-valued spectrograms are obtained by STFT with a 64ms Hamming window and 16ms hop size. The number of channels for the deep complex U-Net is {45, 90, 90, 90, 90, 90, 90, 90, 45, 1}. The convolution kernels are set to (3, 3), and the step sizes are set to (2, 2) except for the middle two layers which are set to (2, 1). The number of TSTBs is 6. In the loss function, we set $\alpha = 0.8$, $\beta = 1/200$, $\gamma = 1$ both in the synthetic and real-world experiments empirically. We use Adam optimizer with a learning rate of 0.001, and use StepLR to adjust the learning rate at an interval of 1 epoch and a multiple of 0.1 times. All experiments are implemented in Pytorch on a NVIDIA GTX1080 Ti GPU.

*5.2. Validation of Only-Noisy Training (ONT) Strategy*

We conduct two experiments to verify the feasibility of our training strategy. Firstly, we conducted an experiment to verify whether our strategy can train the denoising model without clean speech signals. Fig. 4(a) summarizes the evaluated scores, where ONT achieves higher scores than those of input signals (Noisy). The results show that the ONT strategy can train the speech denoising model without clean speech.

Secondly, we use the Mean Opinion Scores (MOS) for subjective evaluation – the new metric provided by the ConferencingSpeech 2021 organizer that is believed to be more correlated with the subjective listening score. Inspired from ITU-T P .835, the evaluation is performed according to the overall quality rating and recorded with 95% confidence intervals (CI). Each tester spends some time familiarizing with the pure clean signal before the evaluation began. The final MOS scores shown in Fig. 4(b) are calculated according to the mean values, ratings of 5 means that the quality is perfect (no artifacts). The results show that our ONT strategy gets the highest MOS scores and shows best speech denoising effect reflected in the subjective evaluation.

*5.3. Comparisons with Other Training Strategies*

Our proposed strategy is evaluated based on the white noise and four kinds of US8K noise. The results are shown in Table 1, and the mean and standard deviation of metrics are calculated separately. The gray-colored rows represent our proposed methods, while values shown in bold denote the best performance achieved by the compared methods.

Furthermore, for a fairer comparison, the NerNT strategy and our ONT strategy are all conducted on the same denoising configuration of NCT and NNT strategy (Kashyap et al., 2021). For NerNT, we selected an additional noise $n$ from US8K, and mixed it to the noisy speech $x$ at randomly selected SNRs between 0 to 10dB. In this experiment, we consider $x$ as the signal and $n$ as noise in SNR. For ONT-rTSTM and ONT-cTSTM, all the encoder and decoder layers have the same configuration as ONT, except that the modules (rTSTM, cTSTM) inserted between them are different. Thus, all methods in Table 1 are conducted with the same amount and variety of training noise samples, and only the amount and variety of target signals are different.

According to the comparative experiments in Table 1, the following conclusions can be drawn:

**Comparison with other strategies:** All the proposed methods (i.e., ONT, ONT-rTSTM, ONT-cTSTM) show superior results compared with other existing strategies (i.e., NCT, NNT, NerNT). Our speech denoising strategy is not only superior to traditional methods training with clean speech, but also exceeds methods with additional noise information. Even in the white noise case where NCT strategy does not exceed NNT, our method also demonstrates the effectiveness and superiority in denoising performance, with each indicator exceeding two benchmark methods.

**Evaluation of cTSTM:** (1) In US8K category 2 (Children Playing), ONT outperforms ONT-cTSTM and ONT-rTSTM, but the differences are negligible. It can be considered that for noisy samples overlaying noises of children playing in the real world, speech features extracted by TSTM has little influence on the denoising network. (2) In US8K category 0 (Air Conditioning), 4 (Drilling), all metrics except STOI in ONT-cTSTM show the best results. STOI is calculated based on the time envelope correlation coefficient of clean speech and noisy speech, which offers a high correlation with speech intelligibility. Therefore, the cTSTM has little effect on improving the speech intelligibility of denoising results in US8K category 0, but improvements on other metrics still work. (3) For white noise and other US8K categories, the adoption of cTSTM is very effective to reconstruct speech features output from the encoder. It makes the denoising network not only process the magnitude and phase information from spectrograms more accurately, but also ensure that the context information is not ignored. In general, our ONT-cTSTM produces superior denoising performance on synthetic and real world noises.

*5.4. Comparisons with State-of-the-Art Methods*

To verify the effectiveness of the model with other state-of-the-art methods, we provide the evaluation results on a benchmark dataset. Except for PESQ, three additional metrics (i.e., CSIG, COVL, CBAK) are also adopted. As can be seen from Table 2, the following conclusions can be drawn:

**Table 1**

Evaluation with other strategies.

| Noise | Model | SNR | SSNR | PESQ-NB | PESQ-WB | STOI |
|---|---|---|---|---|---|---|
| **White** | NCT (Kashyap et al., 2021) | 17.323 ± 3.488 | 4.047 ± 4.738 | 2.655 ± 0.428 | 1.891 ± 0.359 | 0.655 ± 0.17 |
| | NNT (Kashyap et al., 2021) | 16.937 ± 3.973 | 3.752 ± 4.918 | 2.597 ± 0.462 | 1.840 ± 0.375 | 0.650 ± 0.18 |
| | NerNT | - | - | - | - | - |
| | ONT (ours) | 17.563 ± 2.596 | 8.389 ± 2.961 | 2.690 ± 0.347 | 1.878 ± 0.293 | 0.833 ± 0.07 |
| | +rTSTM | 18.137 ± 2.122 | 9.077 ± 2.437 | 2.643 ± 0.317 | **2.003 ± 0.282** | 0.839 ± 0.07 |
| | +cTSTM | **18.209 ± 2.095** | **9.088 ± 2.222** | **2.811 ± 0.288** | 1.997 ± 0.276 | **0.847 ± 0.07** |
| **US8K-0** (Air Conditioning) | NCT (Kashyap et al., 2021) | 4.174 ± 3.608 | −1.433 ± 3.124 | 1.980 ± 0.232 | 1.386 ± 0.165 | 0.578 ± 0.18 |
| | NNT (Kashyap et al., 2021) | 4.656 ± 5.612 | −0.800 ± 3.687 | 2.440 ± 0.386 | 1.658 ± 0.298 | 0.641 ± 0.17 |
| | NerNT | 4.318 ± 4.026 | -1.294 ± 2.188 | 2.140 ± 0.332 | 1.160 ± 0.198 | 0.697 ± 0.18 |
| | ONT (ours) | 6.270 ± 3.711 | 1.185 ± 2.685 | 2.615 ± 0.488 | 1.776 ± 0.283 | **0.900 ± 0.09** |
| | +rTSTM | 6.231 ± 3.773 | 1.314 ± 2.704 | 2.143 ± 0.521 | 1.336 ± 0.300 | 0.809 ± 0.90 |
| | +cTSTM | **6.317 ± 3.813** | **1.414 ± 2.684** | **2.730 ± 0.485** | **1.891 ± 0.294** | 0.806 ± 0.09 |
| **US8K-1** (Car Horn) | NCT (Kashyap et al., 2021) | 4.143 ± 3.899 | −0.415 ± 3.664 | 1.924 ± 0.313 | 1.370 ± 0.208 | 0.562 ± 0.20 |
| | NNT (Kashyap et al., 2021) | 4.823 ± 6.166 | 0.324 ± 4.558 | 2.445 ± 0.481 | 1.770 ± 0.410 | 0.634 ± 0.19 |
| | NerNT | 4.464 ± 3.858 | -0.837 ± 3.714 | 2.121 ± 0.351 | 1.484 ± 0.256 | 0.651 ± 0.19 |
| | ONT (ours) | 6.244 ± 4.039 | 0.382 ± 4.029 | 2.650 ± 0.488 | 1.836 ± 0.324 | 0.759 ± 0.12 |
| | +rTSTM | 6.234 ± 4.051 | 0.505 ± 3.486 | 2.773 ± 0.518 | 1.861 ± 0.286 | 0.761 ± 0.12 |
| | +cTSTM | **6.339 ± 4.045** | **0.609 ± 3.160** | **2.954 ± 0.429** | **1.902 ± 0.376** | **0.850 ± 0.12** |
| **US8K-2** (Children Playing) | NCT (Kashyap et al., 2021) | 3.830 ± 3.580 | −1.403 ± 3.201 | 1.854 ± 0.235 | 1.332 ± 0.152 | 0.550 ± 0.17 |
| | NNT (Kashyap et al., 2021) | 4.348 ± 5.370 | −0.636 ± 3.776 | 2.177 ± 0.378 | 1.512 ± 0.248 | 0.620 ± 0.17 |
| | NerNT | 3.636 ± 3.392 | -1.936 ± 3.103 | 1.812 ± 0.258 | 1.265 ± 0.134 | 0.572 ± 0.17 |
| | ONT (ours) | **6.559 ± 4.440** | **-0.343 ± 3.600** | **3.410 ± 0.504** | **1.963 ± 0.312** | **0.879 ± 0.10** |
| | +rTSTM | 6.546 ± 4.449 | -0.287 ± 3.514 | 3.027 ± 0.502 | 1.806 ± 0.314 | 0.774 ± 0.10 |
| | +cTSTM | 6.442 ± 4.419 | -0.302 ± 3.509 | 3.018 ± 0.503 | 1.867 ± 0.303 | 0.777 ± 0.10 |
| **US8K-3** (Dog Barking) | NCT (Kashyap et al., 2021) | 3.438 ± 3.457 | −0.684 ± 3.767 | 1.773 ± 0.326 | 1.326 ± 0.190 | 0.520 ± 0.18 |
| | NNT (Kashyap et al., 2021) | 3.990 ± 5.451 | −0.002 ± 5.084 | 2.147 ± 0.535 | 1.550 ± 0.372 | 0.593 ± 0.22 |
| | NerNT | 3.537 ± 3.465 | -1.336 ± 3.105 | 1.787 ± 0.260 | 1.249 ± 0.126 | 0.569 ± 0.17 |
| | ONT (ours) | 6.580 ± 6.687 | 3.982 ± 6.959 | 2.181 ± 0.712 | 1.599 ± 0.541 | 0.768 ± 0.17 |
| | +rTSTM | 6.592 ± 6.779 | 4.106 ± 7.386 | 2.193 ± 0.732 | 1.622 ± 0.591 | 0.772 ± 0.18 |
| | +cTSTM | **6.615 ± 6.886** | **4.199 ± 7.390** | **2.193 ± 0.735** | **1.626 ± 0.603** | **0.773 ± 0.17** |

(1) Compared with other strategies which train without the clean target (i.e., NNT, NerNT), our method shows superiority on most metrics (i.e., PESQ, CSIG, COVL). It means that even compared with other novel training strategies, our strategy presents excellent advantages both in training conditions and training effects.

(2) Compared to the network trained with traditional supervised data (i.e., NCT), our method still achieves very competitive performance and outperforms most existing methods, which further verifies the effectiveness of our ONT strategy.

*5.5. Sensitivity Analysis*

In this section, we conduct a series of experiments to investigate the sensitivity of parameters in the ONT strategy.

**Analysis of Sampling Methods**: Here, we analyze the sampling method of the training audio pairs generated module in Fig. 2. In this experiment, we study ONT-cTSTM from two aspects: Influence of the hyper-parameter $k$; Influence of using a random neighbor sub-sampler or a fixed location sub-sampler.

We conduct six comparison experiments by combining three values of $k$ and two kinds of sub-samplers. In each sampling area (size $k$), all time domain values are from the corresponding location of the same sub-sampled audio. Then two sub-sampled audio signals are randomly selected from the original sampled audio or selected from the fixed location in each sampling area.

The evaluation results of PESQ-NB are shown in Table 3. It can be seen from the results that when the sampling interval is 2 and the sampling method is "Random", the speech denoising effect performs best. We further conclude that this sampling interval could represent the similarity between the sub-sampled pair best. Additionally, by comparing the two sampling methods, it's obvious that the random neighbor sub-sampler outperforms the fixed location sub-sampler. Due to the importance of randomness in the sampling strategy, the fixed location sub-sampler is only a special case of random sub-sampler, so the random sub-sampler gets better results.

**Analysis of Regularization Loss**: Next, we analyze the regularization loss while training the denoising module in (13). In this experiment, we select 6

### Table 2
Comparisons with other methods on VoiceBank-DEMAND.

| Strategy | Model | PESQ | CSIG | COVL | CBAK |
|---|---|---|---|---|---|
| | Noisy | 1.97 | 3.35 | 2.63 | 2.44 |
| | SEGAN (Pascual et al., 2017) | 2.16 | 3.48 | 2.80 | 2.94 |
| | MMSE-GAN (Soni et al., 2018) | 2.53 | 3.80 | 3.12 | 3.14 |
| | SERGAN (Baby and Verhulst, 2019) | 2.62 | - | - | - |
| NCT | BLSTM (MSE) (Weninger et al., 2015) | 2.71 | 3.94 | 3.28 | 3.32 |
| | MetricGAN (Fu et al., 2019) | 2.86 | 3.99 | 3.42 | 3.18 |
| | HiFi-GAN (Su et al., 2020) | 2.94 | 4.07 | 3.07 | 3.49 |
| | PHASEN (Yin et al., 2020) | 2.99 | 4.21 | 3.62 | 3.55 |
| | CAUNet (Wang et al., 2021) | 2.96 | 4.22 | 3.60 | 3.53 |
| | T-GSA (Kim et al., 2020) | 3.06 | 4.18 | 3.62 | **3.59** |
| | DEMUCS (Defossez et al., 2020) | 3.07 | **4.31** | 3.63 | 3.40 |
| | MetricGAN+ (Fu et al., 2021) | **3.15** | 4.14 | **3.64** | 3.16 |
| NNT | NeTT (Fujimura et al., 2021) | *2.63* | *3.77* | *3.19* | *3.37* |
| NerNT | NyTT(L) (Fujimura et al., 2021) | *2.31* | *3.23* | *2.75* | *3.02* |
| ONT | **ONT-cTSTM(ours)** | ***3.12*** | ***4.02*** | ***3.55*** | ***2.98*** |

values (0, 1, 2, 4, 10, 20, 40) for experiments on ONT-cTSTM.

The evaluation results of PESQ-NB are shown in Table 4, which shows that when $\gamma$ is greater than 1, it will bring a worse effect. This is because a larger $\gamma$ will make the basic loss less important while training, thus leading to a worse denoising effect. However, while $\gamma$ is zero, there is no regularization loss while training, and the training process would not address the essential difference of the ground-truth values between the sub-sampled pairs. Therefore, it's inappropriate and would lead to over-smoothing while training. We further analyze the best performance while $\gamma$ is 1. The basic loss and regularization loss achieve an optimal solution, which not only avoids over-smoothing, but also achieves better denoising effect.

**Analysis of complex-TSTM**: To verify that the cTSTM is helpful for performance improvement, we conduct studies on ONT by adding different TSTMs, which consist of multiple TSTBs.

We experiment with three different numbers of TSTBs (4, 6, 8) in real or complex way. For example, 4rTSTB means there is a real TSTM between the encoder and decoder layers, and the TSTM consists of four TSTBs. In all comparison models, each encoder and decoder layer have the same configuration as ONT.

The experimental results are shown in Table 5, which shows that 6cTSTB configuration brings the best performance, and when the number of TSTBs is larger or smaller, it will bring a worse effect. This is because that the encoder and decoder layers in ONT focus on extracting both the magnitude and phase features efficiently, while the TSTB module pays attention on the local and global context information. To this end, the number of TSTBs acts as a controller between these two denoising effects. From another perspective, almost all complex models from the table outperform real models obviously, which means that our proposed cTSTM is very effective for improving the performance of speech denoising.

### 5.6. Visualization Results

To visually see the effectiveness of the ONT strategy, an utterance of a clean speech signal as well as its noisy version and its denoised version are exhibited in Table 6. As can be seen from the noisy signals in Table 6(a) and Table 6(d), after overlapping white noise and real world noise, there is a multitude of noise interferences. From the Table 6(c) and Table 6(f), we can observe that our proposed method effectively has a good noise removal

### Table 3
Ablation on different sampling methods.

| Noise | k=2 | | k=4 | | k=6 | |
|---|---|---|---|---|---|---|
| | Fixed | Random | Fixed | Random | Fixed | Random |
| **White** | 2.694 | **2.811** | 2.670 | 2.677 | 2.533 | 2.607 |
| **US8K-0** | 2.599 | **2.730** | 2.368 | 2.492 | 2.334 | 2.484 |
| **US8K-1** | 2.803 | **2.954** | 2.612 | 2.745 | 2.310 | 2.373 |
| **US8K-2** | 2.595 | **3.018** | 2.433 | 2.481 | 2.390 | 2.397 |
| **US8K-3** | 1.986 | **2.193** | 1.931 | 1.955 | 1.921 | 1.933 |

### Table 4
Ablation on different weights of regularization loss.

| Noise | $\gamma$=0 | $\gamma$=1 | $\gamma$=2 | $\gamma$=4 | $\gamma$=10 | $\gamma$=20 | $\gamma$=40 |
|---|---|---|---|---|---|---|---|
| **White** | 2.762 | **2.811** | 2.798 | 2.795 | 2.799 | 2.801 | 2.793 |
| **US8K-0** | 2.652 | **2.730** | 2.705 | 2.701 | 2.681 | 2.677 | 2.671 |
| **US8K-1** | 2.879 | **2.954** | 2.936 | 2.890 | 2.892 | 2.888 | 2.881 |
| **US8K-2** | 2.982 | **3.018** | 3.012 | 3.002 | 2.998 | 2.992 | 2.980 |
| **US8K-3** | 1.999 | **2.193** | 2.111 | 2.107 | 2.084 | 2.003 | 1.989 |

### Table 5
Ablation on different network configuration.

| Noise | 4rTSTB | 4cTSTB | 6rTSTB | 6cTSTB | 8rTSTB | 8cTSTB |
|---|---|---|---|---|---|---|
| **White** | 2.717 | 2.723 | 2.643 | **2.811** | 2.793 | 2.809 |
| **US8K-0** | 2.156 | 2.633 | 2.143 | **2.730** | 2.566 | 2.603 |
| **US8K-1** | 2.699 | 2.818 | 2.773 | **2.954** | 2.740 | 2.897 |
| **US8K-2** | 2.986 | 3.009 | **3.027** | 3.018 | 2.879 | 3.010 |
| **US8K-3** | 2.189 | 2.187 | 2.193 | **2.193** | 2.177 | 2.179 |

**Table 6**

Example of speech time-frequency diagram.

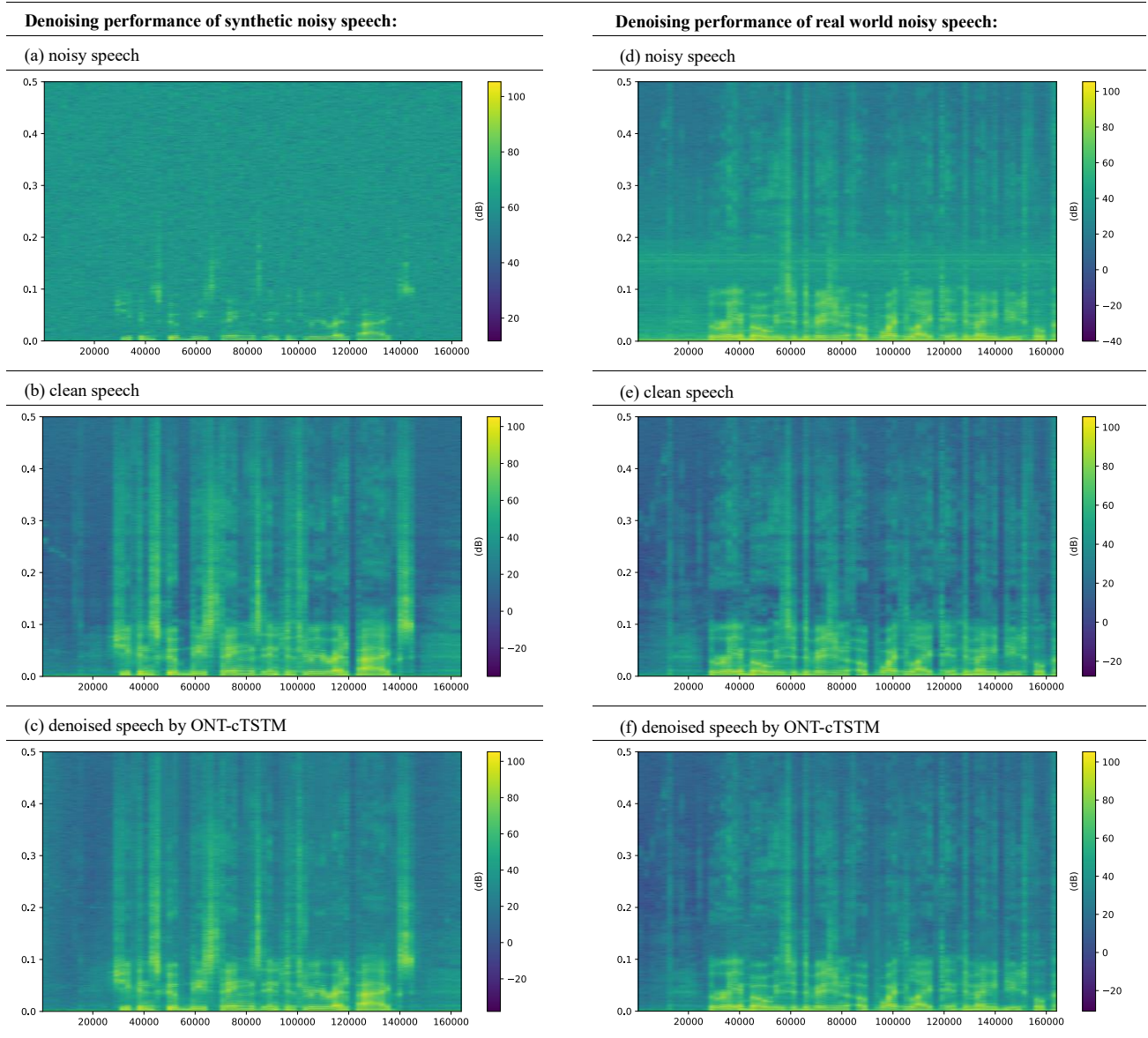

performance. It retains more speech harmonic segments while the speech distortion was minimized, which demonstrates the effectiveness of our method.

## 6. Conclusion

In this paper, we propose a novel self-supervised speech denoising strategy, which overcomes the limitation of clean speech acquisition and puts an end to the need for additional costly data. Our approach successfully solves the Only-Noisy Training problem by generating sub-sampled paired signals with audio sub-samplers, and realizes an effective complex-valued speech denoising network for better denoising performance. Experimental results confirm that our proposed strategy outperforms other compared strategies for most evaluation metrics. Furthermore, even compared to the state-of-the-art methods, our strategy still shows competitiveness in scenarios where clean speech is limited and rare. In the future, we would like to extend the proposed method to scenes of multimodal denoising, in order to highlight the advantages of the Only-Noisy speech denoising method in multi-task scenarios where audio samples are limited and rare.

**CRediT authorship contribution statement**

**Jiasong Wu:** Conceptualization, Methodology, Writing- Reviewing and Editing. **Qingchun Li:** Writing- Original draft preparation, Software, Visualization. **Guanyu Yang:** Validation, Project administration. **Lei Li:** Data Curation, Validation. **Lotfi Senhadji:** Formal analysis, Writing- Reviewing and Editing. **Huazhong Shu:** Supervision, Resources.

**Declaration of competing interest**

The authors declare that they have no known competing financial interests or personal relationships that could have appeared to influence the work reported in this paper.

**Acknowledgements**

This work was supported in part by the National Key Research and Development Program of China (No. 2021ZD0113202), and in part by the National Natural Science Foundation of China under Grants 61876037, 62171125, 31800825, 61871117, 61871124, 61773117, 61872079, and in part by the Pre-Research Foundation of 50912040302, and in part by INSERM under the Grant the calls IAL and IRP. The authors would like to thank the Big Data Computing Center of Southeast University for providing the facility support on the numerical calculations in this paper.